%% file: TVT25.tex
\documentclass[conference, 10pt]{IEEEtran}
\IEEEoverridecommandlockouts
\usepackage{cite}
\usepackage{amsmath,amssymb,amsfonts}
\usepackage{graphicx}
\usepackage{textcomp}
\usepackage{xcolor}
\usepackage{subcaption}
\usepackage{bm}
\usepackage{multirow}
\usepackage{booktabs}
\usepackage{subcaption}
\usepackage{amsthm}\usepackage{balance}\usepackage{url}
\usepackage{algorithm,algpseudocode}
\usepackage{amssymb}
\usepackage{eqnarray}

\newtheorem*{proof*}{Proof}

\algrenewcommand\algorithmicrequire{\textbf{Initialization:}}
\algrenewcommand\algorithmicensure{\textbf{Output:}}
\algdef{SE}[DOWHILE]{Do}{doWhile}{\algorithmicdo}[1]{\algorithmicwhile\ #1}%

\makeatletter

\newcommand{\Rmnum}[1]{\expandafter\@slowromancap\romannumeral #1@}
\makeatother

\date{}

\def\BibTeX{{\rm B\kern-.05em{\sc i\kern-.025em b}\kern-.08em
    T\kern-.1667em\lower.7ex\hbox{E}\kern-.125emX}}

\begin{document}
\title{Robust MIMO Semantic Communication with Imperfect CSI via Knowledge Distillation
}


\author{Mingze Gong,~\IEEEmembership{Graduate Student Member,~IEEE}, 
    Shuoyao Wang,~\IEEEmembership{Senior Member,~IEEE},\\
    Shijian Gao,~\IEEEmembership{Member,~IEEE},
    Jia Yan,~\IEEEmembership{Member,~IEEE},
    and Suzhi Bi~\IEEEmembership{Senior Member,~IEEE}
    }

\maketitle

\input{abstract.tex}
\input{introduction.tex}

\input{system_model.tex}

\input{proposed_methods.tex}

\input{simulation_results.tex}

\input{conclusion.tex}

\bibliography{ref}    
\bibliographystyle{ieeetr}
\end{document}

%% file: abstract.tex
\begin{abstract}\label{abstract}
    Semantic communication (SemComm) has emerged as a new communication paradigm.
    To enhance efficiency, multiple-input-multiple-output (MIMO) technology has been further integrated into SemComm systems.
    However, existing MIMO SemComm systems assume perfect channel matrix estimation for channel-adaptive joint source-channel coding, which is impractical due to hardware and pilot overhead constraints. 
    In this paper, we propose a semantic image transmission system with channel matrix and channel noise adaptation, named HANA-JSCC, to cope with channel estimation errors in MIMO systems. 
    We propose a channel matrix adaptor that collaborates with the channel codec to adapt to misaligned channel state information, thereby mitigating the impact of estimation errors.
    Since the relationship between the estimated channel matrix and true channel matrix is ill-posed (one-to-many), we further introduce a two-stage training strategy with knowledge distillation to overcome the convergence difficulties caused by the ill-posed problem.
    Comparing with the state-of-the-art benchmarks, HANA-JSCC achieves $0.40\sim0.54$dB higher average performance across various noise and estimation error levels in various datasets. 

\end{abstract}
\begin{IEEEkeywords}
    Semantic communication, MIMO, imperfect channel estimation, channel adaptation, image transmission. 
\end{IEEEkeywords}

%% file: introduction.tex
\section{Introduction} \label{introduction}
    \IEEEPARstart{s}{ixth} generation (6G) has been conceptualized as an intelligent information system both being driving and driven by deep learning (DL) technology \cite{letaief2019roadmap}. 
    By leveraging the advances in DL, semantic communication (SemComm) systems typically use deep neural networks (DNNs) for joint source-channel coding (JSCC) to extract and encode the semantic information and can transmit various types of sources, such as images\cite{bourtsoulatze2019deep}.
    Beyond, some works \cite{zhang2023predictive, 10589474} have developed channel-aware JSCC to to improve noise robustness. 
    For instance, \cite{10589474} leveraged attention mechanism to adapt to various channel conditions based on signal-noise ratio (SNR). 

    Most existing SemComm research focuses on single-input-single-output (SISO) communications, while multiple-input-multiple-output (MIMO) technique, widely adopted in modern devices, offers significant benefits in data rates and system capacity. 
    Motivated by this, recently, some works \cite{10293784, 10713884, 10681856} have investigated the combination of conventional SemComm and MIMO technique.
    In particular, \cite{10713884} has exploited the SNR difference of decomposed equivalent SISO sub-channels to achieve high-fidelity transmission through singular value decomposition (SVD) precoding and post-processing. 
    
    Inspired by SNR-adaptive JSCC schemes, some works \cite{10510413, 10680080, 10559783, 10597355} have utilized the feedback channel state information (CSI) in MIMO systems to enhance the robustness to fading effects during transmission. 
    For instance, \cite{10510413} proposed to integrate attention mechanism with channel matrix into semantic encoding to achieve adaptation to fading channels in MIMO systems. 

    Integrating the advanced MIMO SemComm systems with large-scale antenna array holds great potential for improving communication efficiency\cite{10680080}.
    However, existing works rely on perfectly estimated CSI for robustness enhancement towards noisy channels.
    While channel estimation is relatively straightforward in SISO systems (i.e., SNR estimation), accurate acquisition of full channel matrices in MIMO systems is both costly and challenging due to constraints on pilot overhead and hardware resources, especially in large-scale MIMO scenarios \cite{10379539}.
    Alternatively, we show in the preliminary experiments that the imperfect channel estimation might significantly degrades SemComm performance.
    Although~\cite{10648086} avoids estimation overhead via pilot-free MIMO, the channel-independent coding limits transmission performance.
    Overall, achieving efficient SemComm in MIMO antenna array systems with imperfect channel matrices remains an open problem.

    Inspired by above, we develop a semantic image transmission system with channel matrix and channel noise adaptation, named HANA-JSCC, for MIMO communications with channel estimation errors. 
    To address the challenges posed by inaccurate CSI, we propose refining the compressed latent semantics during channel coding to enhance robustness against different channel realizations.
    To this end, we develop a channel matrix adaptation module into channel codec. 
    To address the convergence challenges arising from the ill-posed one-to-many mapping between the estimated and true channel matrices, a two-stage training strategy leveraging knowledge distillation is employed.
    The major contributions of this paper are summarized as follows:
    \begin{itemize}
        \item A MIMO SemComm system is developed for image transmission with misaligned CSI feedback. 
        To the best of our knowledge, we are among the first to consider the SemComm in MIMO systems over imperfect channel estimation.

        \item We propose a channel matrix adaptor with an adaptation mechanism and optimization strategy to reduce performance loss from misaligned channel matrix.
        It refines compressed features within the channel codecs and is boosted via a two-stage strategy using knowledge distillation from a perfect-CSI teacher model.        

        \item Compared with the representative benchmarks, HANA-JSCC achieves remarkable performance under various datasets. 
        On average, HANA-JSCC displays $0.40 \sim 0.54$dB more PSNR than the state-of-the-art methods. 

    \end{itemize}

%% file: system_model.tex
\section{System Model and Problem Description} \label{system_model}
\begin{figure}[!t]
        \centering
        \includegraphics[scale=0.442]{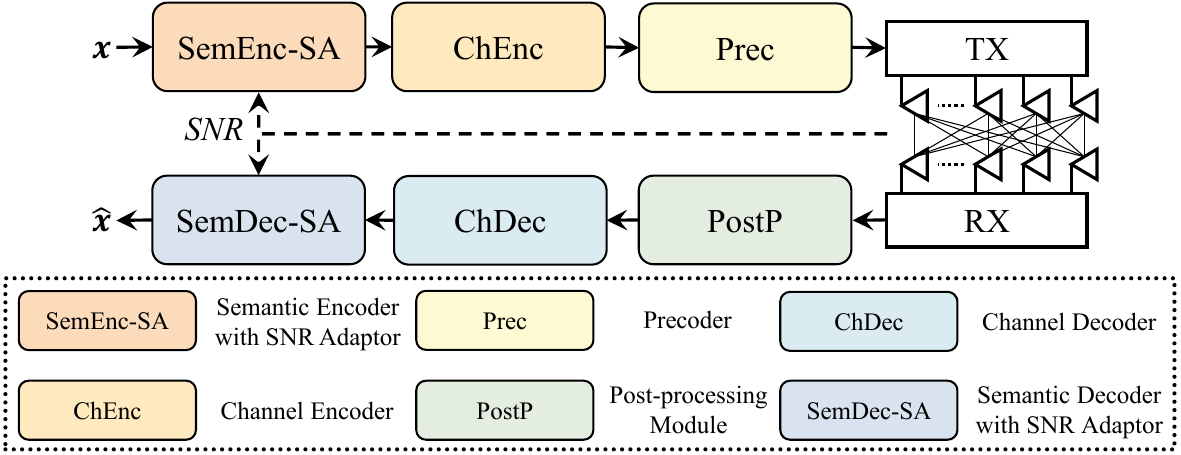}
        \caption{\small{System model of the SNR-adaptive MIMO SemComm system.}}
        \label{fig:system_model}
    \end{figure}

    In this section, we introduce the general system model of the MIMO SemComm system for image transmission.
    Furthermore, we discuss the formulated problem.
    As shown in Fig.~\ref{fig:system_model}, we consider a single-user SNR-adaptive MIMO SemComm system for image transmission. 
    Notably, the transmitter and receiver are equipped with $n_{\text{tx}}$ and $n_{\text{rx}}$ antennas. 
    Without loss of generality, the input of the MIMO SemComm system is an image $\bm{x} \in \mathbb{R}^{c \times h \times w}$ with the number of color channels $c$ as well as the height $h$ and the width $w$. 
    The output is reconstructed image $\hat{\bm{x}} \in \mathbb{R}^{c \times h \times w}$. 


\subsection{System Model of MIMO SemComm}
\subsubsection{Transmitter}
    The transmitter contains a semantic encoder with an SNR-adaptation module, a channel encoder, and a precoder. 
    The semantic encoder extracts features through learned function $\text{E}_\mathrm{S} :: \mathbb{R}^{c \times h \times w} \to \mathbb{R}^{c' \times h' \times w'}$, where $c', h'$ and $w'$ represent dimension, height, and width of the extracted feature map.
    Following semantic encoder, a channel encoder is deployed to compress semantic features for communication efficiency. 
    The function of channel encoding is denoted as $\text{E}_\mathrm{C} :: \mathbb{R}^{c' \times h' \times w'} \to \mathbb{C}^{n_\mathrm{tx} \times d}$, where $d$ is the number of transmitted symbols in each antenna. 
    Briefly, the semantic and channel encoding functions are given by 
    \begin{align}
        \bm{z}_\mathrm{S} &= \text{E}_\mathrm{S}(\bm{x}, r | \bm{\phi}_\mathrm{S}), \label{semantic_encoding} \\
        \bm{z}_\mathrm{C} &= \text{E}_\mathrm{C}(\bm{z}_\mathrm{S} | \bm{\phi}_\mathrm{C}), \label{channel_encoding}
    \end{align}
    where $\bm{z}_\mathrm{S} \in \mathbb{R}^{c' \times h' \times w'}$ is the encoded semantic features and $r$ is the SNR of the noisy channel. 
    Furthermore, $\bm{z}_\mathrm{C} \in \mathbb{C}^{n_\mathrm{tx} \times d}$ is the compressed semantic features. 
    $\bm{\phi}_\mathrm{S}$ and $\bm{\phi}_\mathrm{C}$ are the sets of learnable parameters of semantic encoder and channel encoder, respectively. 
    Notably, a non-trainable power normalization module is incorporated at the end of the channel encoder.
        
    Indicated by \cite{10713884}, SVD precoding and post-processing benefit MIMO SemComm systems via channel decomposition. 
    Inspired by this, we consider SVD precoding process in MIMO SemComm system. 
    Thus, given the channel matrix $\bm{H} \in \mathbb{C}^{n_{\text{rx}} \times n_{\text{tx}}}$, the SVD precoding process can be given by 
    \begin{equation}
        \bm{H} = \bm{U}\bm{\Sigma}\bm{V}^\mathrm{H}, 
    \end{equation}
    where $\bm{U} \in \mathbb{C}^{n_{\text{rx}} \times n_{\text{rx}}}$ and $\bm{V} \in \mathbb{C}^{n_{\text{tx}} \times n_{\text{tx}}}$ are the unitary matrices. 
    Notably, $\bm{V}^\mathrm{H}$ is a conjugate transpose matrix of $\bm{V}$. 
    Additionally, $\bm{\Sigma} \in \mathbb{C}^{n_{\text{rx}} \times n_{\text{tx}}}$ contains the non-negative singular values as diagonal elements, whose total number is subjected to $\text{min}(n_{\text{tx}}, n_{\text{rx}})$. 
    Wihout loss of generality, the singular values are listed in descending order. 
    Accodingly, the precoded signal $\bm{z} \in \mathbb{C}^{n_{\text{tx}} \times d}$ can be represented as 
    \begin{equation}
        \bm{z} = \bm{V}\bm{z}_\mathrm{C}. 
    \end{equation}
    Then, the precoded signal is transmitted to the receiver through MIMO channel. 

\subsubsection{Physical Channel}
    Following \cite{10713884}, we consider a flat Rayleigh fading channel between transceivers in this paper.
    The received signal $\hat{\bm{z}} \in \mathbb{C}^{n_{\text{rx}} \times d}$ at the receiver is given by 
    \begin{equation}
        \hat{\bm{z}} = \bm{H}\bm{z} + \bm{N} = \bm{U}\bm{\Sigma}\bm{V}^\mathrm{H}\bm{V}\bm{z}_\mathrm{C} + \bm{N} = \bm{U}\bm{\Sigma}\bm{z}_\mathrm{C} + \bm{N},
    \end{equation}
    where $\bm{N} \in \mathbb{C}^{n_{\text{rx}} \times d}$ represents the additive white Gaussian noise (AWGN). 
    In addition, $\bm{N}$ follows the complex Gaussian distirbution $\mathcal{CN}(0, \sigma^2_\mathrm{n})$, where $\sigma^2_\mathrm{n}$ dentoes the noise variance.  
        
\subsubsection{Receiver}
    Similar with the transmitter, the receiver consists of a post-processing module, a channel decoder and a semantic decoder with SNR-adaptation. 
    Given SVD precoding, the received signal is equalized by 
    \begin{equation}
        \hat{\bm{z}}_\mathrm{C} = \bm{U}^\mathrm{H}\hat{\bm{z}} = \bm{U}^\mathrm{H}\bm{U}\bm{\Sigma}\bm{z}_\mathrm{C} + \bm{U}^\mathrm{H}\bm{N} = \bm{\Sigma}\bm{z}_\mathrm{C} + \bm{U}^\mathrm{H}\bm{N}, 
    \end{equation}
    where $\hat{\bm{z}}_\mathrm{C} \in \mathbb{C}^{n_{\text{rx}} \times d}$ is the equalized signal. 
    Accordingly, the inference result $\bm{y}$ is given by 
    \begin{align}
        \hat{\bm{z}}_\mathrm{S} &= \text{D}_\mathrm{C}(\hat{\bm{z}}_\mathrm{C} | \bm{\theta}_\mathrm{C}), \label{channel_decoding} \\
        \hat{\bm{x}} &= \text{D}_\mathrm{S}(\hat{\bm{z}}_\mathrm{S} | \bm{\theta}_\mathrm{S}), \label{semantic_decoding}
    \end{align}
    where $\hat{\bm{z}}_\mathrm{S} \in \mathbb{R}^{c' \times h' \times w'}$ is the recovered semantic features. 
    Respectively, $\text{D}_\mathrm{C} \in \mathbb{C}^{n_{\text{rx}} \times d} \to \mathbb{R}^{c' \times h' \times w'}$ and $\text{D}_\mathrm{S} \in \mathbb{R}^{c' \times h' \times w'} \to \mathbb{R}^{c \times h \times w}$ are the channel and semantic decoding functions with learnable parameters $\bm{\theta}_\mathrm{C}$ and $\bm{\theta}_\mathrm{S}$.

\subsection{Problem Description}
    As discussed in Section~\ref{introduction}, existing works face significant performance degradation due to inaccurate channel estimation.    
    In practice, perfect CSI, especially $\bm{H}$, is costly and hard to obtain, given the large number of parameters to estimate and feedback~\cite{1624653}.
    It becomes even more challenging when the MIMO antenna array shifts from small-scale to large-scale (e.g. from $4\times4$ to $16\times16$)\cite{10379539}.
    According to \cite{1624653}, the estimated channel matrix $\bm{H}_\mathrm{est} \in \mathbb{C}^{n_{\text{rx}} \times n_{\text{tx}}}$ can be given by 
    \begin{equation}
        \bm{H}_\mathrm{est} = \bm{H}_\mathrm{p} + \bm{H}_\mathrm{e},
        \label{estimation_error}
    \end{equation}
    where $\bm{H}_\mathrm{p} \in \mathbb{C}^{n_{\text{rx}} \times n_{\text{tx}}}$ is the actual channel matrix. 
    Moreover, $\bm{H}_\mathrm{e} \in \mathbb{C}^{n_{\text{rx}} \times n_{\text{tx}}}$ is the estimation error and it follows the complex Gaussian distribution $\mathcal{CN}(0, \sigma^2_\mathrm{e})$. 
    Therefore, the MIMO transmission over physical channel can be rewritten as 
    \begin{equation}
        \hat{\bm{z}} = \bm{H}_\mathrm{p}\bm{V}_\mathrm{est}\bm{z}_\mathrm{C} + \bm{N} = \bm{U}_\mathrm{p}\bm{\Sigma}_\mathrm{p}\bm{V}^\mathrm{H}_\mathrm{p}\bm{V}_\mathrm{est}\bm{z}_\mathrm{C} + \bm{N}.
        \label{imp_transmission}
    \end{equation}
    Similarly, the post-processing can be rewritten as 
    \begin{equation}
        \tilde{\bm{z}} = \bm{U}^\mathrm{H}_\mathrm{est}\hat{\bm{z}} = \bm{U}^\mathrm{H}_\mathrm{est}\bm{U}_\mathrm{p}\bm{\Sigma}_\mathrm{p}\bm{V}^\mathrm{H}_\mathrm{p}\bm{V}_\mathrm{est}\bm{z}_\mathrm{C} + \bm{U}^\mathrm{H}_\mathrm{est}\bm{N}.
        \label{imp_postp}
    \end{equation}
    
    Eq.~(\ref{estimation_error}) indicates that a single $\bm{H}_\mathrm{est}$ can correspond to multiple realizations of $\bm{H}_\mathrm{e}$ and $\bm{H}_\mathrm{p}$, despite the shared distribution. 
    The ill-posed one-to-many mapping causes misalignment that confuses the encoder and decoder neural networks during training, hindering convergence.
    Additionally, (\ref{imp_transmission}) and (\ref{imp_postp}) reveal that estimation errors disrupt the decomposition of MIMO channels into parallel SISO sub-channels, ultimately limiting the transmission efficiency of MIMO SemComm systems, as evidenced in \cite{10713884}.

    \begin{figure}[!t]
        \centering
        \includegraphics[scale=0.37]{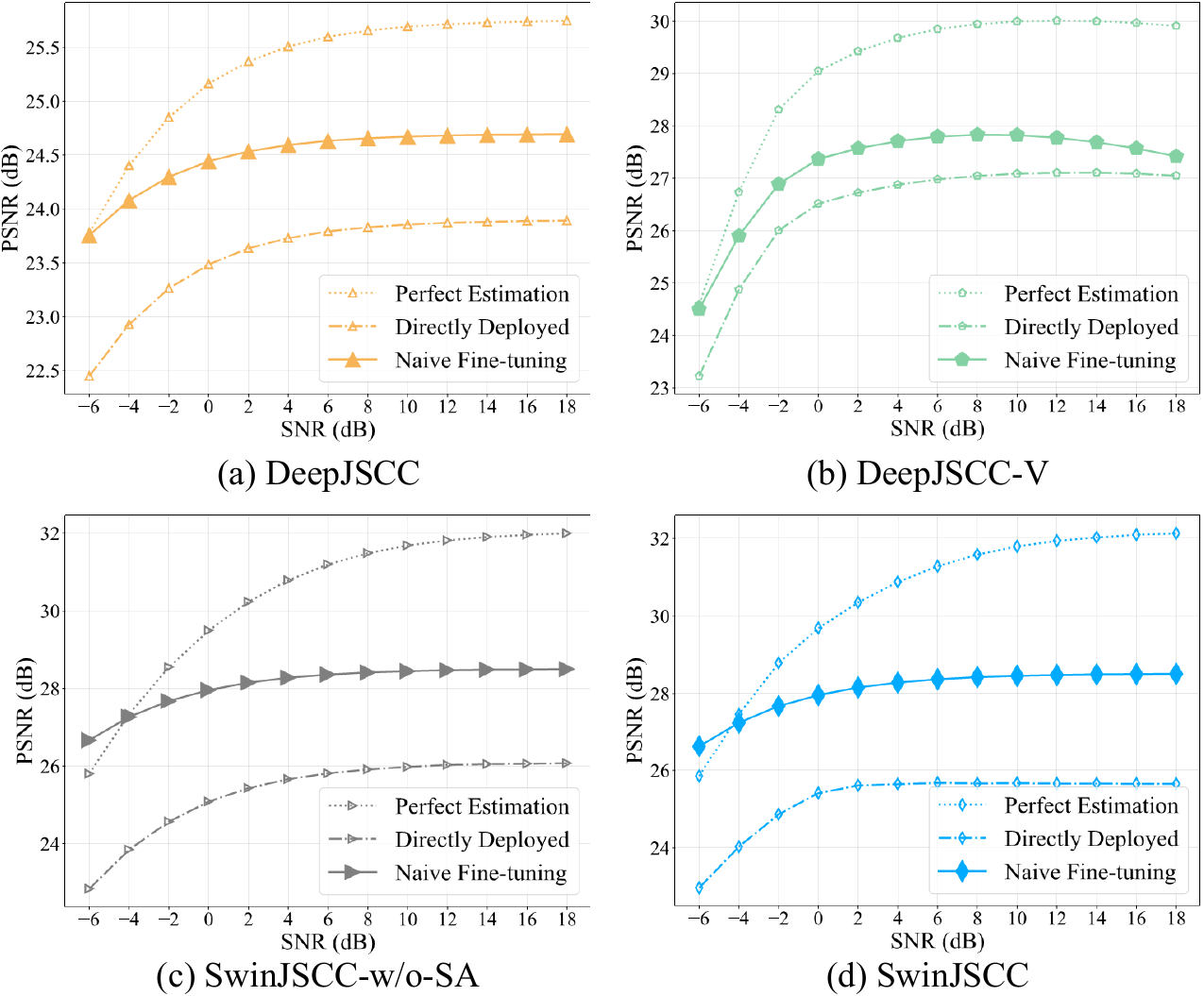}
        \caption{\small{Simulation results on representative SemComm systems in MIMO communications versus SNR of the addtive noise. }}
        \label{fig:problem_description}
    \end{figure}

\subsection{Preliminary Experiment}
    Although MIMO SemComm systems can benefit from fine-tuning with $\bm{H}_\mathrm{est}$, significant performance degradation remains.
    In this subsection, we further experimentally validate this phenomenon in Fig.~\ref{fig:problem_description}.

    \subsubsection{Simulation Settings}
        Without loss of generality, we investigate image transmission with MIMO antenna array using DeepJSCC~\cite{bourtsoulatze2019deep}, DeepJSCC-V~\cite{zhang2023predictive}, SwinJSCC~\cite{10589474}, and SwinJSCC-w/o-SA, which are representative SemComm systems.
        The systems are evaluated under three conditions: 
        \begin{itemize}
         \item Perfect Estimation: SemComm systems are trained and evaluated with $\bm{H}_\mathrm{p}$. 
         \item Directly Deployed: SemComm systems are trained with $\bm{H}_\mathrm{p}$ and evaluated with $\bm{H}_\mathrm{est}$. 
         \item Naive Fine-tuning: SemComm systems are trained and evaluated with $\bm{H}_\mathrm{est}$.
        \end{itemize}
        Following prior works \cite{bourtsoulatze2019deep, zhang2023predictive, 10589474}, the metric to evaluate transmission performance is peak signal-noise ratio (PSNR). 
        Detailed system and training settings can be found in Section~\ref{simulation_results}.~A. 

        \subsubsection{Observation}
        It can be observed in Fig.~\ref{fig:problem_description} that directly deploying systems trained with accurate channel matrix into imperfect scenarios results in significant performance degradation.
        Although simple fine-tuning with misaligned CSI improves performance, the degradation remains substantial.
        Respectively, for DeepJSCC-V and SwinJSCC, the ``Naive Fine-tuning'' approach leads to on average $1.82$ dB and $2.36$ dB lower PSNR compared to ``Perfect Estimation'' method.

    Accordingly, \emph{MIMO SemComm for image transmission with channel estimation errors calls for further investigation. }

%% file: proposed_methods.tex
\section{Methodology and Network Architecture} \label{methodology}
    To address this challenge, we propose HANA-JSCC for robust semantic image transmission.
    In the following, we detail its methodology and network architecture.

\subsection{Methodology}

    \begin{figure}[!t]
        \centering
        \includegraphics[scale=0.34]{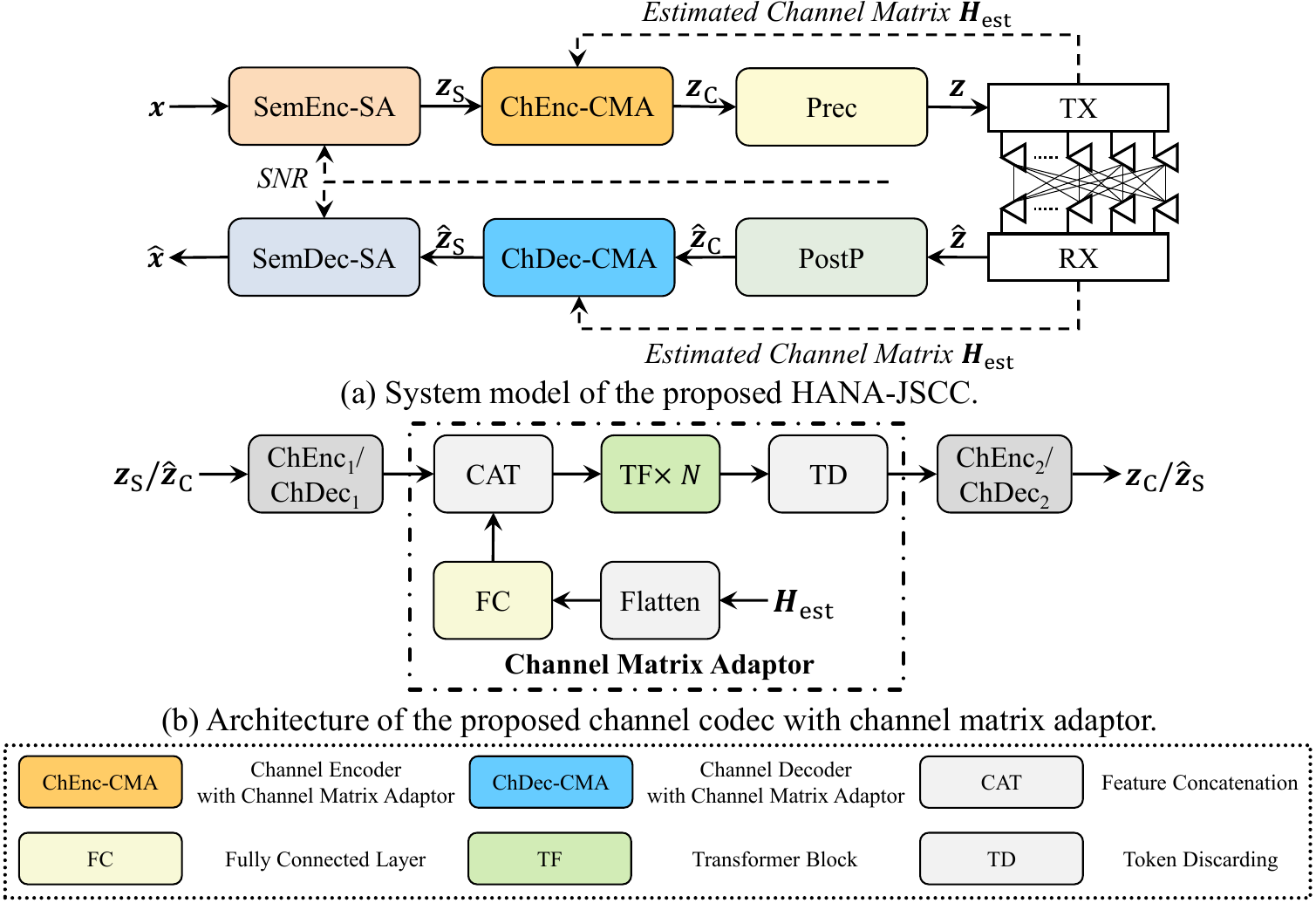}
        \caption{\small{Architecture of the proposed HANA-JSCC.}}
        \label{fig:proposed_method}
    \end{figure}

    Inspired by the advantages of DL-based SemComm \cite{bourtsoulatze2019deep}, HANA-JSCC is constructed based on DNNs. 
    As shown in Fig.~\ref{fig:proposed_method}(a), HANA-JSCC includes an SNR-adaptive semantic encoder and decoder, a channel codec with the proposed channel matrix adaptor, and a precoder/post-processing module.
    
    We propose feature refinement within the channel codecs, avoiding the inefficiencies of both pre- and post-enhancement to better preserve semantic information under misaligned channels.
    Specifically, we divide the channel codec into two sub-codecs and integrate the channel matrix adaptor between them.
    In the following, we rewrite the channel encoding and decoding functions in (\ref{channel_encoding}) and (\ref{channel_decoding}), respectively.

\subsubsection{Channel Encoding with Channel Matrix Adaptation}
    The channel encoder in (\ref{channel_encoding}) is divided into two sub-encoders. 
    The output semantic feature $\bm{z}_\mathrm{C,1} \in \mathbb{R}^{n_\mathrm{tx} \times d'}$ of the first sub-encoder is given by 
    \begin{equation}
        \bm{z}_\mathrm{C,1} = \text{E}_\mathrm{C,1}(\bm{z}_\mathrm{S} | \bm{\phi}_{C,1}),
        \label{zc1}
    \end{equation}
    where $\text{E}_\mathrm{C,1} :: \mathbb{R}^{c' \times h' \times w'} \to \mathbb{R}^{n_\mathrm{tx} \times d'}$ is the compression function with learnable parameters $\bm{\phi}_\mathrm{C,1}$.
    As shown in Fig.~\ref{fig:proposed_method}(b), the channel matrix adaptor follows the first channel sub-encoder.
    Given the semantic feature $\bm{z}_\mathrm{C,1}$ and estimated channel matrix $\bm{H}_\mathrm{est}$, the output feature $\bm{z}_\mathrm{C,2} \in \mathbb{R}^{n_\mathrm{tx} \times d'}$ of channel matrix adaptor can be written as 
    \begin{equation}
        \bm{z}_\mathrm{C,2} = \text{E}_\mathrm{CMA}(\bm{z}_\mathrm{C,1}, \bm{H}_\mathrm{est} | \bm{\phi}_\mathrm{CMA}), 
        \label{zc2}
    \end{equation}
    where $\text{E}_\mathrm{CMA} :: \mathbb{R}^{n_\mathrm{tx} \times d'} \to \mathbb{R}^{n_\mathrm{tx} \times d'}$ is the adaptation function of channel matrix adaptor in transmitter with learnable parameters $\bm{\phi}_\mathrm{CGA}$. 
    Subsequently, the second channel sub-encoder takes $\bm{z}_\mathrm{C,2}$ as input and further compresses the semantic features following
    \begin{equation}
        \bm{z}_\mathrm{C} = \text{E}_\mathrm{C,2}(\bm{z}_\mathrm{C,2} | \bm{\phi}_{C,2}),
        \label{zc}
    \end{equation}
    where $\text{E}_\mathrm{C,2} :: \mathbb{R}^{n_\mathrm{tx} \times d'} \to \mathbb{C}^{n_\mathrm{tx} \times d}$ is the compression function with learnable parameters $\bm{\phi}_\mathrm{C,2}$.

\subsubsection{Channel Decoding with Channel Matrix Adaptation}  
    Similar with the channel encoding process with channel matrix adaptor, the proposed channel decoding can be given by
    \begin{align}
        \hat{\bm{z}}_\mathrm{C,1} &= \text{D}_\mathrm{C,1}(\hat{\bm{z}}_\mathrm{C} | \bm{\theta}_\mathrm{C,1}), \label{hat_zc1}\\
        \hat{\bm{z}}_\mathrm{C,2} &= \text{D}_\mathrm{CMA}(\hat{\bm{z}}_\mathrm{C,1}, \bm{H}_\mathrm{est} | \bm{\theta}_\mathrm{CMA}), \label{hat_zc2}\\
        \hat{\bm{z}}_\mathrm{S} &= \text{D}_\mathrm{C,2}(\hat{\bm{z}}_\mathrm{C,2} | \bm{\theta}_\mathrm{C,2}),\label{hat_zs}
    \end{align}
    where $\hat{\bm{z}}_\mathrm{C,1} \in \mathbb{R}^{n_\mathrm{rx} \times d'}$ and $\hat{\bm{z}}_\mathrm{C,2} \in \mathbb{R}^{n_\mathrm{rx} \times d'}$ are the latent features. 
    Moreover, $\text{D}_\mathrm{C,1} :: \mathbb{C}^{n_\mathrm{rx} \times d} \to \mathbb{R}^{n_\mathrm{rx} \times d'}$ and $\text{D}_\mathrm{C,2} :: \mathbb{R}^{n_\mathrm{rx} \times d'} \to \mathbb{R}^{c' \times h' \times w'}$ are the reconstructed functions with learnbale parameters $\bm{\theta}_\mathrm{C,1}$ and $\bm{\theta}_\mathrm{C,2}$, respectively. 
    Additionally, $\text{D}_\mathrm{CMA} :: \mathbb{R}^{n_\mathrm{rx} \times d'} \to \mathbb{R}^{n_\mathrm{rx} \times d'}$ is the function of channel matrix adaptor in receiver with learnable parameters $\bm{\theta}_\mathrm{CGA}$. 

\subsection{Neural Network Design}
    Indicated by \cite{10589474}, Swin-Transformer based SemComm systems outperform convolutional neural network based ones like DeepJSCC \cite{bourtsoulatze2019deep}. 
    Motivated by this, HANA-JSCC is built upon SwinJSCC\cite{10589474}, employing Swin-Transformer and Channel-ModNet \cite{10589474} for backbone and SNR-adaptation module, with fully connected layers (FCs) based channel sub-codecs for feature compression and recovery. 

    In this paper, we propose to consider the transmitted symbols in each antenna as a token, which is inspired by \emph{the natural data structure of the MIMO transmission}. 
    Moreover, we leverage Transformer to mitigate the performance degradation caused by channel estimation errors.
    

    As shown in Fig.~\ref{fig:proposed_method}(b), the channel matrix adaptor uses an FC to project $\bm{H}_\mathrm{est}$ into a CSI token, which is prepended to the feature sequence from the first channel sub-codec.
    Then, $N$ Transformer blocks enhance feature robustness to estimation errors.
    The CSI token is discarded at the end, and the refined features are passed to the second sub-codec.

\subsection{Two-Stage Training Strategy with Knowledge Distillation}
    In this paper, we propose a two-stage training strategy to boost the proposed channel matrix adaptor. 
    In Stage-I, we adopt a general end-to-end training process with $\bm{H}_\mathrm{est}$. 
    To accelerate convergence and improve performance, we initialize the model parameters with those of SwinJSCC pretrained under imperfect channel estimation conditions.
    Subsequently, the system is trained with imperfect channel matrix in (\ref{estimation_error}), where $\sigma^2_\mathrm{e}$ is sampled from a uniform distribution $\mathcal{U}(\sigma^2_\mathrm{e, min}, \sigma^2_\mathrm{e, max})$. 
    The loss function in Stage-I is given by 
    \begin{equation}
        \mathcal{L}_\mathrm{S_1} = \Vert \hat{\bm{x}} - \bm{x} \Vert_1. 
    \end{equation}
    To alleviate the burden of fitting the channel distribution under channel estimation errors in MIMO systems, we freeze the network parameters of semantic codecs.
    Accordingly, only the channel codecs with channel matrix adaptor are trained. 
    
    As indicated by Section~\ref{system_model}.~C, MIMO SemComm systems benefit from fine-tuning process with $\bm{H}_\mathrm{est}$, referring to Stage-I. 
    However, as the system adapts to misaligned conditions modeled in (\ref{estimation_error}), the performance under near-perfect estimation deteriorates.
    Motivated by this, we leverage knowledge distillation to further improve transmission performance Stage-II.
    Specifically, the well-trained system in Stage-I serves as the student model, with all parameters tunable to fully exploit the optimization potenital for performance improvement.
    Meanwhile, a system with perfect channel knowledge in Stage-I acts as the teacher model, sharing the same network architecture but with parameters frozen.
    We consider Kullback-Leibler (KL) divergence \cite{10345474} as the knowledge distillation loss. 
    Accordingly, the loss for Stage-II training is given by 
    \begin{equation}
        \mathcal{L} = \mathcal{L}_\mathrm{S_1} + \beta (\text{D}_\mathrm{KL}(\bm{z}_\mathrm{C, St} || \bm{z}_\mathrm{C, Tc}) + \text{D}_\mathrm{KL}(\hat{\bm{z}}_\mathrm{S, St} || \hat{\bm{z}}_\mathrm{S, Tc})),
        \label{loss_s2}
    \end{equation}
    where $\text{D}_\mathrm{KL}(\cdot)$ is the KL divergence and $\beta$ is a weight hyperparameter. 
    Moreover, $\bm{z}_\mathrm{C, i}$ and $\hat{\bm{z}}_\mathrm{S, i}$ are the semantic features accessed through (\ref{zc1})-(\ref{zc}) and (\ref{hat_zc1})-(\ref{hat_zs}), respectively, where $i \in \{\mathrm{Tc}, \mathrm{St}\}$. 
    Notably, $\mathrm{Tc}$ and $\mathrm{St}$ denote the teacher and student models, respectively.

%% file: simulation_results.tex
\section{Simulation Results} \label{simulation_results}
    In this section, we evaluate the image transmission performance of the proposed HANA-JSCC and benchmarks. 

\subsection{Simulation Settings}
    \subsubsection{Dataset and System Setups}
        Simulations are conducted on datasets in different scales, including CIFAR10 and Kodak24. 
        CIFAR10 dataset consists 50,000 R.G.B images at the size of $3\times32\times32$ for training and 10,000 images for test. 
        To further evaluate HANA-JSCC, we introduce Kodak24 dataset, which includes 24 R.G.B. images at the size of $3 \times 512 \times 768$ or $3 \times 768 \times 512$. 
        Without loss of generality, we consider large-scale MIMO antenna arrays at both the transmitter and receiver to evaluate the proposed system and benchmarks under inevitable channel estimation errors arising from resource-constrained conditions.
        Particularly, the number of transceivers antenna $n_\mathrm{tx}$ and $n_\mathrm{rx}$ are both set to 16, aligning with \cite{10510413}. 
        We adopt the SwinJSCC's\footnote{https://github.com/semcomm/SwinJSCC} settings~\cite{10589474} and adopt 6 Transformer blocks in the channel matrix adaptor.
        Moreover, the compression ratio is set to $1/12$ for CIFAR10 and $1/48$ for Kodak24.

    \subsubsection{Benchmarks}

        To validate the superiority of HANA-JSCC, we consider the following five representative benchmarks: DeepJSCC~\cite{bourtsoulatze2019deep}, DeepJSCC-V~\cite{zhang2023predictive}, DeepJSCC-MIMO~\cite{10597355}, SwinJSCC~\cite{10589474}, and SwinJSCC-w/o-SA. 
        All benchmarks are pre-trained with accurate CSI and then fine-tuned using inaccuratly estimated channel matrix.

    \subsubsection{Training Settings}
        During training, noise levels of $\bm{N}$ are randomly chosen from $[1, 3, 5, 7, 9]$, and $\sigma^2_\mathrm{e}$ is sampled from $\mathcal{U}(0.01, 0.1)$, which is common in practice\cite{6891254}.
        The loss weight $\beta$ is set to 1, and the batch size in Stage-II is double that in Stage-I.
        For CIFAR10, systems are trained and tested with batch size 256.
        For Kodak24, we follow~\cite{zhang2023predictive} using 48,627 ImageNet validation images for training, with $3 \times 256 \times 256$ crops and batch size 16.
        In addition, we adopt Adam as the optimizer with learning rate (LR) of $1\times10^{-3}$ and cosine annealing strategy to dynamicly adjust LR.

    \begin{figure}[!t]
        \centering
        \begin{subfigure}[t]{0.48\columnwidth}
            \centering
            \includegraphics[width=\linewidth]{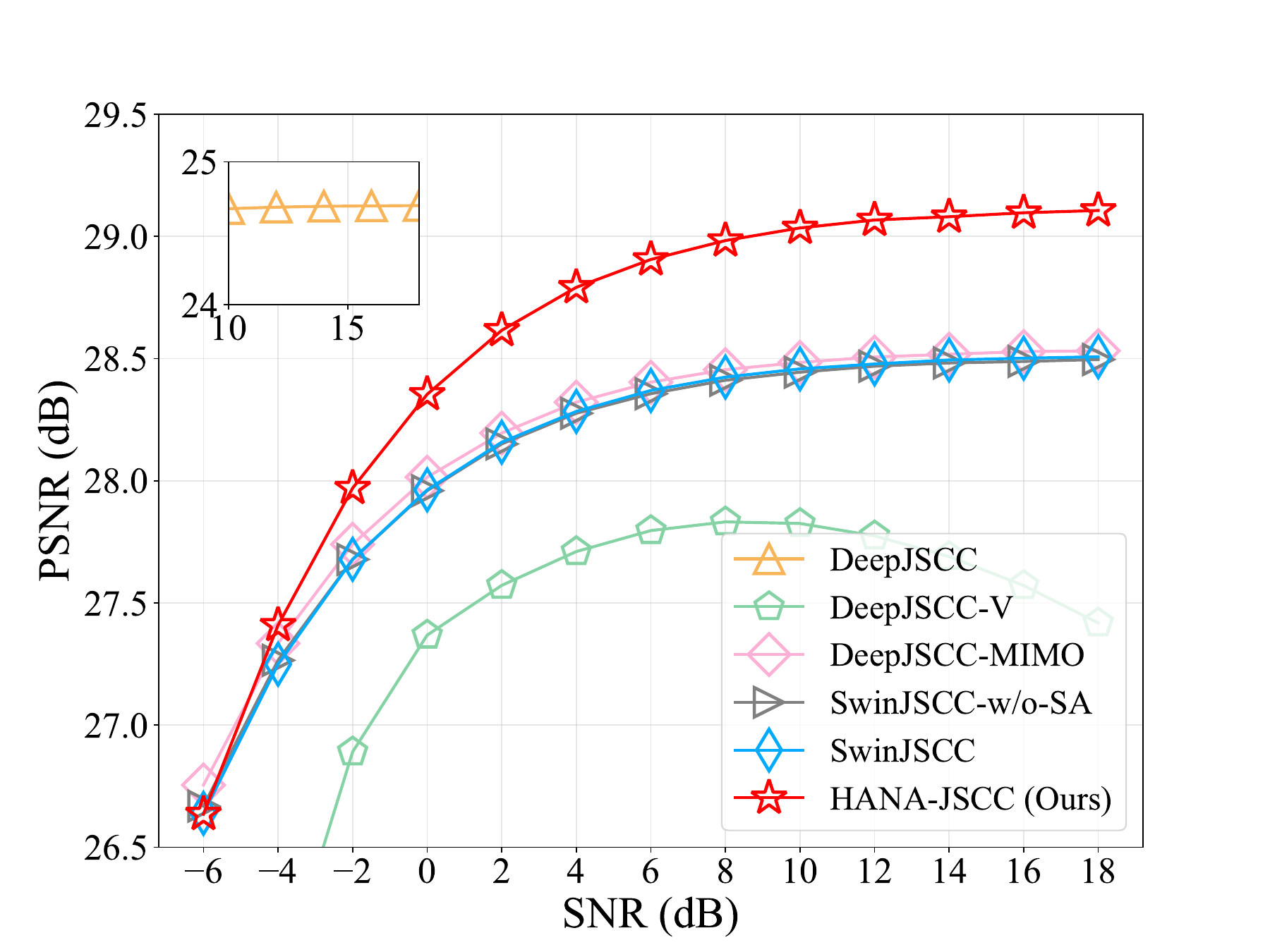}
            \caption{PSNR vs. SNR}
            \label{fig:psnr_vs_N}
        \end{subfigure}
        \hfill
        \begin{subfigure}[t]{0.48\columnwidth}
            \centering
            \includegraphics[width=\linewidth]{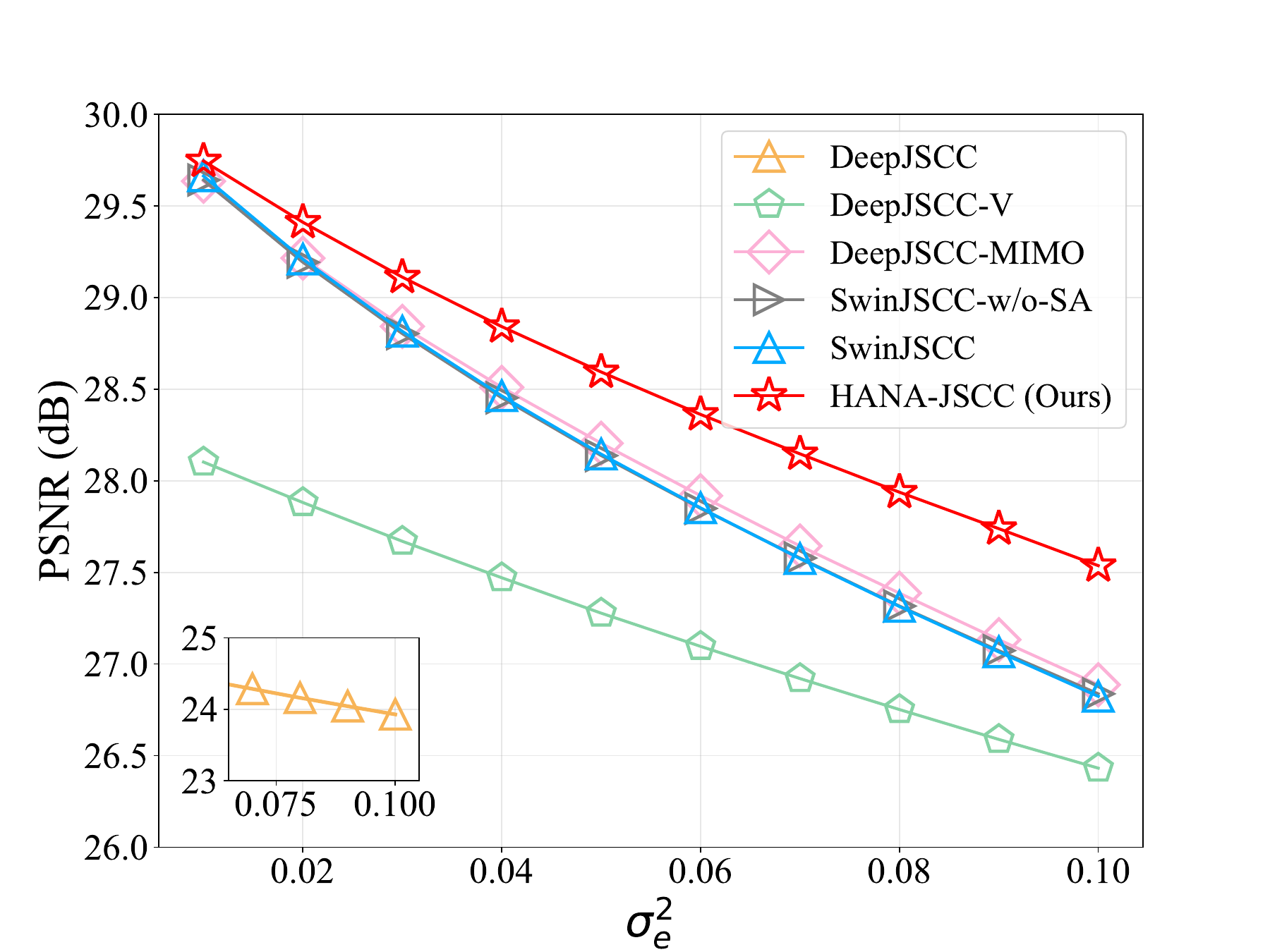}
            \caption{PSNR vs. $\sigma_\mathrm{e}$}
            \label{fig:psnr_vs_H}
        \end{subfigure}
        \caption{\small{The PSNR of MIMO SemComm systems versus SNR from -6dB to 18dB and $\sigma_\mathrm{e}$ from 0.01 to 0.1 in CIFAR10 dataset. }}
        \label{fig:simulation_results_cifar10}
    \end{figure}

\subsection{Comparisons}
    \subsubsection{Simulation Results on CIFAR10}        
        In this experiment, we evaluate the transmission performance of the proposed HANA-JSCC and benchmark systems over MIMO channels across a wide range of SNRs and estimation errors on the CIFAR10 dataset.
        In Fig.~\ref{fig:simulation_results_cifar10}, we compare the achieved PSNR across all six approaches. 
        Overall, HANA-JSCC displays more performance across all SNRs and estimation errors, showcasing its superiority in misaligned channel conditions. 
        On average, HANA-JSCC achieves a PSNR of $28.54$ dB for image reconstruction, outperforming DeepJSCC-MIMO and SwinJSCC by $0.40$ dB and $0.45$ dB, respectively.
    

    \begin{figure}[!t]
        \centering
        \begin{subfigure}[t]{0.48\columnwidth}
            \centering
            \includegraphics[width=\linewidth]{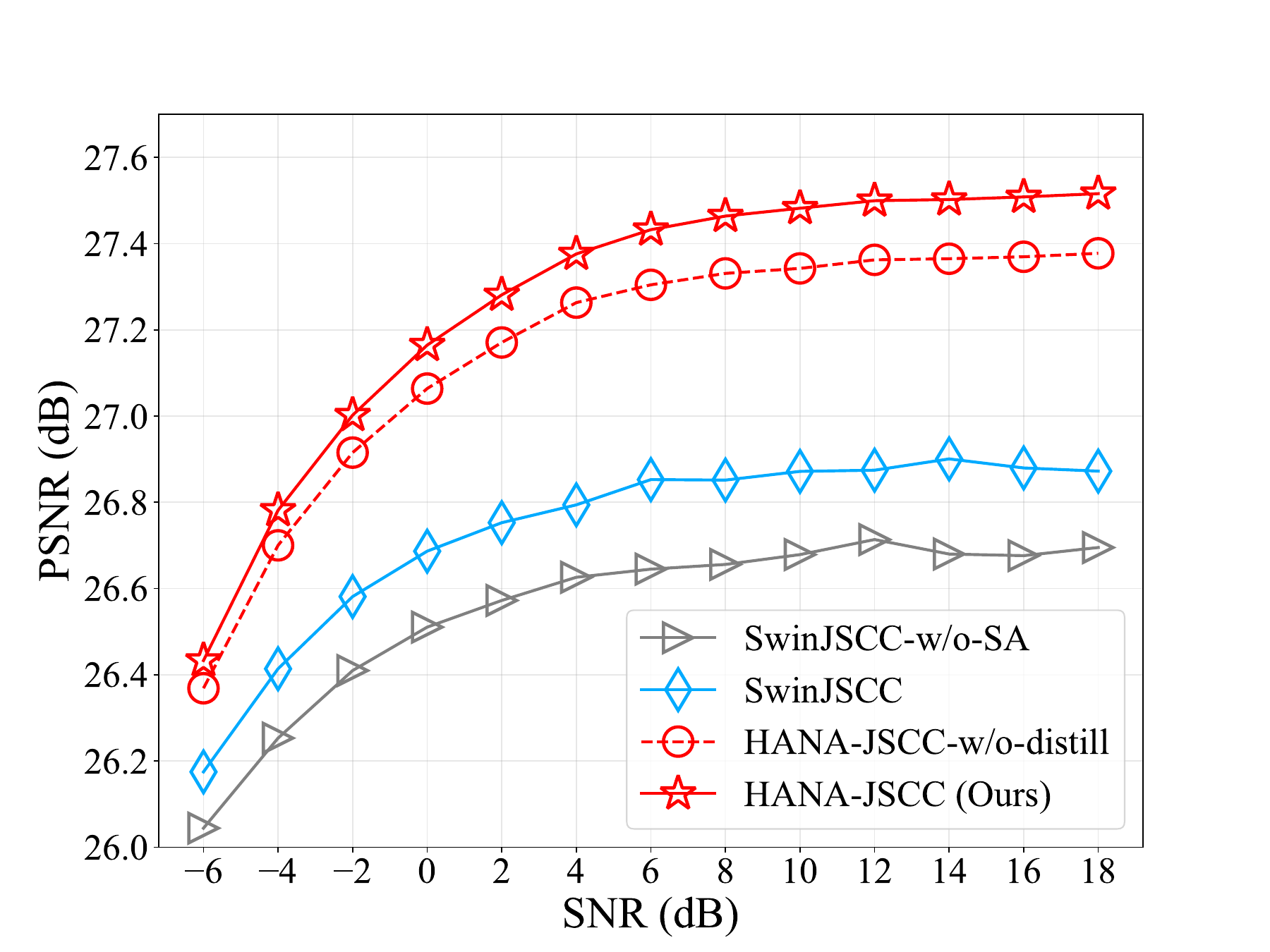}
            \caption{PSNR vs. SNR}
            \label{fig:psnr_vs_N_kodak24}
        \end{subfigure}
        \hfill
        \begin{subfigure}[t]{0.48\columnwidth}
            \centering
            \includegraphics[width=\linewidth]{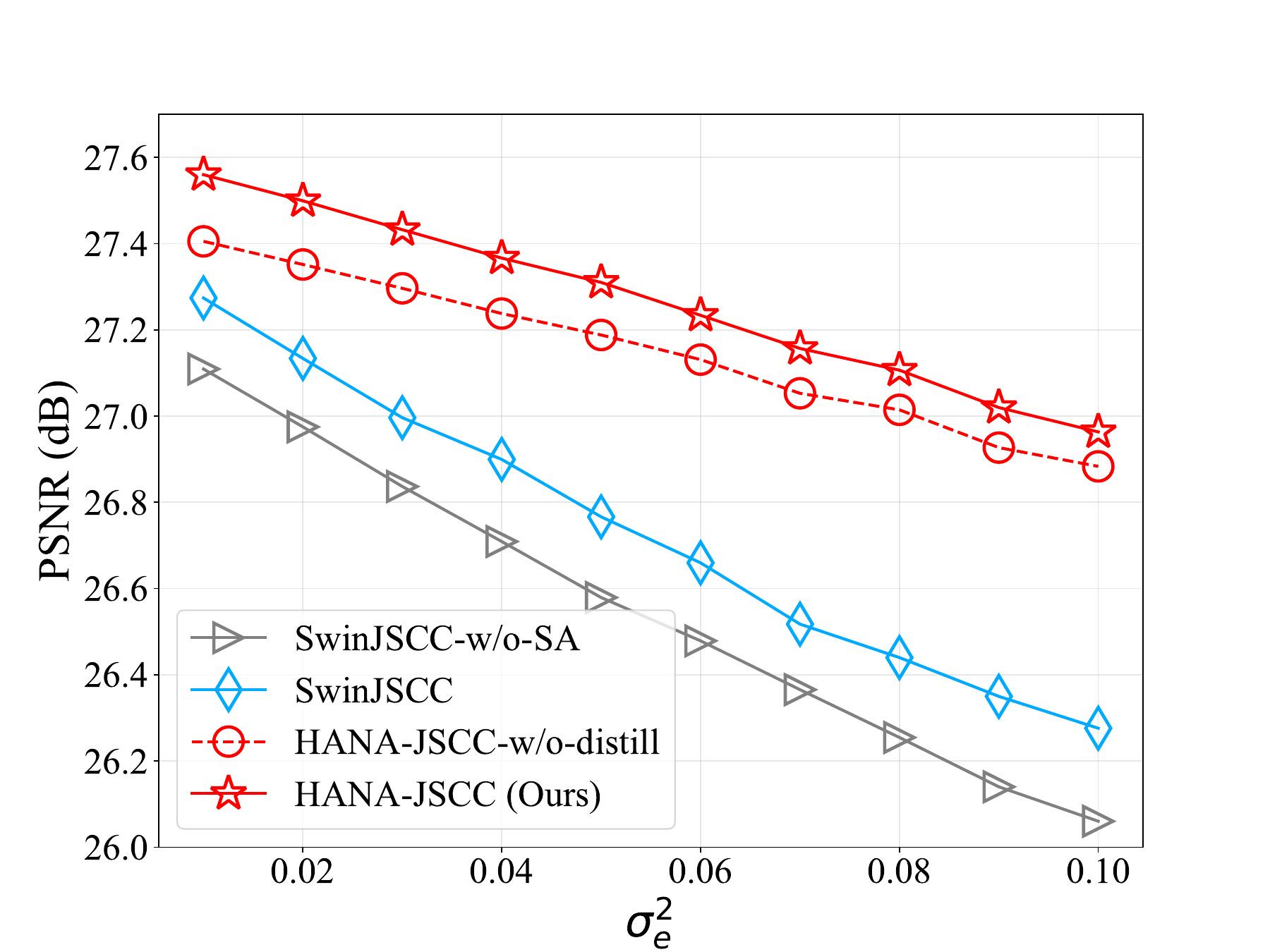}
            \caption{PSNR vs. $\sigma_\mathrm{e}$}
            \label{fig:psnr_vs_H_kodak24}
        \end{subfigure}
        \caption{\small{The PSNR of MIMO SemComm systems versus SNR from -6dB to 18dB and $\sigma_\mathrm{e}$ from 0.01 to 0.1 in Kodak24 dataset. }}
        \label{fig:simulation_results_kodak24}
    \end{figure}

    \subsubsection{Simulation Results on Kodak24}
        To further evaluate the effectiveness of the proposed component and training strategy, we conduct simulations on the high-resolution Kodak24 dataset.
        To streamline the simulations, we restrict the following evaluations to Swin-Transformer-based SemComm systems. 
        Moreover, we consider the proposed method without knowledge distillation as another baseline, named HANA-JSCC-w/o-distill. 
        Simulation results are shown in Fig.~\ref{fig:simulation_results_kodak24}. 
        Overall, HANA-JSCC maintains superior performance in image transmission over MIMO communications with imperfect CSI.
        Specifically, HANA-JSCC-w/o-distill achieves an average PSNR gain of 0.42 dB over SwinJSCC, highlighting the benefits of the proposed channel matrix adaptor. 
        Additionally, HANA-JSCC outperforms its non-distilled counterpart by 0.12 dB on average, demonstrating the effectiveness of the knowledge distillation-based training strategy.

%% file: conclusion.tex
\section{Conclusion}\label{conclusion}
    In this paper, we propose a SemComm system, named HANA-JSCC, to address channel estimation errors in MIMO-based image transmission.
    To mitigate performance degradation from imperfect channel estimation, we propose a channel matrix adaptor with an adaptation mechanism and optimization strategy.
    The adaptor refines compressed semantic features through a channel matrix adaptation module integrated into the channel codec. 
    Given the ill-posed (one-to-many) nature of the relationship between estimated and true channel matrices, we further introduce a two-stage training strategy based on knowledge distillation to facilitate stable convergence.
    Simulation results have illustrated the remarkable performance of the proposed HANA-JSCC. 
    Particularly, HANA-JSCC achieves $0.40\sim0.54$dB higher average performance across various SNRs and estimation errors in various datasets.